\documentclass[twocolumn,aps,english,prl,nofootinbib,floatfix]{revtex4-1}
\usepackage{mathrsfs}
\usepackage{graphicx}
\usepackage{amsmath, amssymb}
\usepackage{babel}
\usepackage{color}
\usepackage{slashed}
\usepackage{hyperref}
\usepackage[export]{adjustbox}
\definecolor{orange}{RGB}{255,127,0}

\begin{document}

%%%%%%%%%%%%%%%%%%%%%%%%%%%%%%%%%%%%%%%%%%%%%%%%%%%%%%%%%%%%%%%%%%%%%
%%%%%%%%%%%%%%%%%%%%%%%%%%%%%%%%%%%%%%%%%%%%%%%%%%%%%%%%%%%%%%%%%%%%%
\title{Reduced uncertainty of the axial $\gamma Z$-box correction to the proton's weak charge}

\author{Jens Erler$^{a,b}$}
\author{Mikhail Gorchtein$^{c}$}
\author{Oleksandr Koshchii$^{a}$}
\author{Chien-Yeah Seng$^{d}$}
\author{Hubert Spiesberger$^{a,e}$}

\affiliation{$^{a}$PRISMA$^+$Cluster of Excellence,
    Institut f\"ur Physik,\\
	Johannes Gutenberg-Universit\"at, D-55099 Mainz, Germany}
\affiliation{$^{b}$Departamento de F\'{\i}sica Te\'{o}rica,
    Instituto de F\'{\i}sica,\\
	Universidad Nacional Aut\'{o}noma de M\'{e}xico, 04510 CDMX,
	M\'{e}xico}
\affiliation{$^{c}$PRISMA$^+$Cluster of Excellence,
    Institut f\"ur Kernphysik,\\
	Johannes Gutenberg-Universit\"at, D-55099 Mainz, Germany}
\affiliation{$^{d}$Helmholtz-Institut f\"ur Strahlen- und Kernphysik
    and Bethe Center for Theoretical Physics,\\ Universit\"at Bonn,
    53115 Bonn, Germany}
\affiliation{$^{e}$Centre for Theoretical and Mathematical Physics
    and Department of Physics,\\
	University of Cape Town, Rondebosch 7700, South Africa}

\date{\today}

%%%%%%%%%%%%%%%%%%%%%%%%%%%%%%%%%%%%%%%%%%%%%%%%%%%%%%%%%%%%%%%%%%%%%
\begin{abstract}
We present the fully up-to-date calculation of the $\gamma Z$-box 
correction which needs to be taken into account to
determine the weak mixing angle at low energies from
parity-violating electron proton scattering. We make use of
neutrino and antineutrino inclusive scattering data to predict
the parity-violating structure function $F_3^{\gamma Z}$ by
isospin symmetry. 
%This allows us to reduce the uncertainty in
%the $\gamma Z$-box obtained from that structure function. 
Our
new analysis confirms previous results for the axial contribution
to the $\gamma Z$-box graph, and reduces the uncertainty by a
factor of~2. In addition, we note that the presence of
parity-violating photon-hadron interactions induces an additional
contribution {\em via\/} $F_3^{\gamma \gamma}$. Using experimental and
theoretical constraints on the nucleon anapole moment we are able
to estimate the uncertainty associated with this contribution. We point out 
that future measurements are expected to significantly reduce this latter uncertainty.
\end{abstract}

\maketitle

%%%%%%%%%%%%%%%%%%%%%%%%%%%%%%%%%%%%%%%%%%%%%%%%%%%%%%%%%%%%%%%%%%%%%
\section{Introduction}
\label{sec:intro}

The precision measurement of $s_W^2 \equiv
\sin^2\theta_W$, where $\theta_W$ is the Standard Model (SM)
weak mixing angle, in parity-violating (PV) electron scattering
serves as a powerful tool to test the SM and
search for physics beyond it. Since the
energy dependence of $s_W^2$ is very precisely
predicted within the SM~\cite{Erler:2017knj}, any significant deviation from it
would be an indication of Beyond Standard Model
(BSM) physics. Polarized elastic $ep$ scattering
in the limit of vanishing beam energy $E$ and
momentum transfer $t$, probes the so-called weak charge of the proton,
\begin{equation}
Q_W^p = - \lim_{|t|\rightarrow 0}
\left.\frac{4\sqrt{2}\pi\alpha}{G_F |t|}
\frac{d\sigma_+ - d\sigma_-}{d\sigma_+ + d\sigma_-}\right|_{E=0}.
\end{equation}
Here, $\alpha$ is the fine structure constant and $G_F$ the
Fermi constant. $d\sigma_{\pm}$ are the differential cross
sections for scattering with right-handed and left-handed polarized electrons,
respectively. At tree-level this quantity is given by $1-4s_W^2$
which is accidentally suppressed. This effectively leads to an
enhancement in the sensitivity to $s_W^2$,
\begin{equation}
\frac{\Delta s_W^2}{s_W^2}
\approx 0.08\, \frac{\Delta Q_W^p}{Q_W^p},
\end{equation}
which also implies an enhanced sensitivity to BSM effects that may
enter the running of $s_W^2$. As a concrete example, the upcoming
P2 experiment at the Mainz Energy-Recovering Superconducting Accelerator (MESA), that plans for a measurement of the
proton weak charge to 1.4\% with a beam energy of $E=155$~MeV will
lead to a determination of $s_W^2$ to 0.15\% precision~\cite{Becker:2018ggl}.

\begin{figure}[t]
	\begin{centering}
		\includegraphics[width=0.7\linewidth]{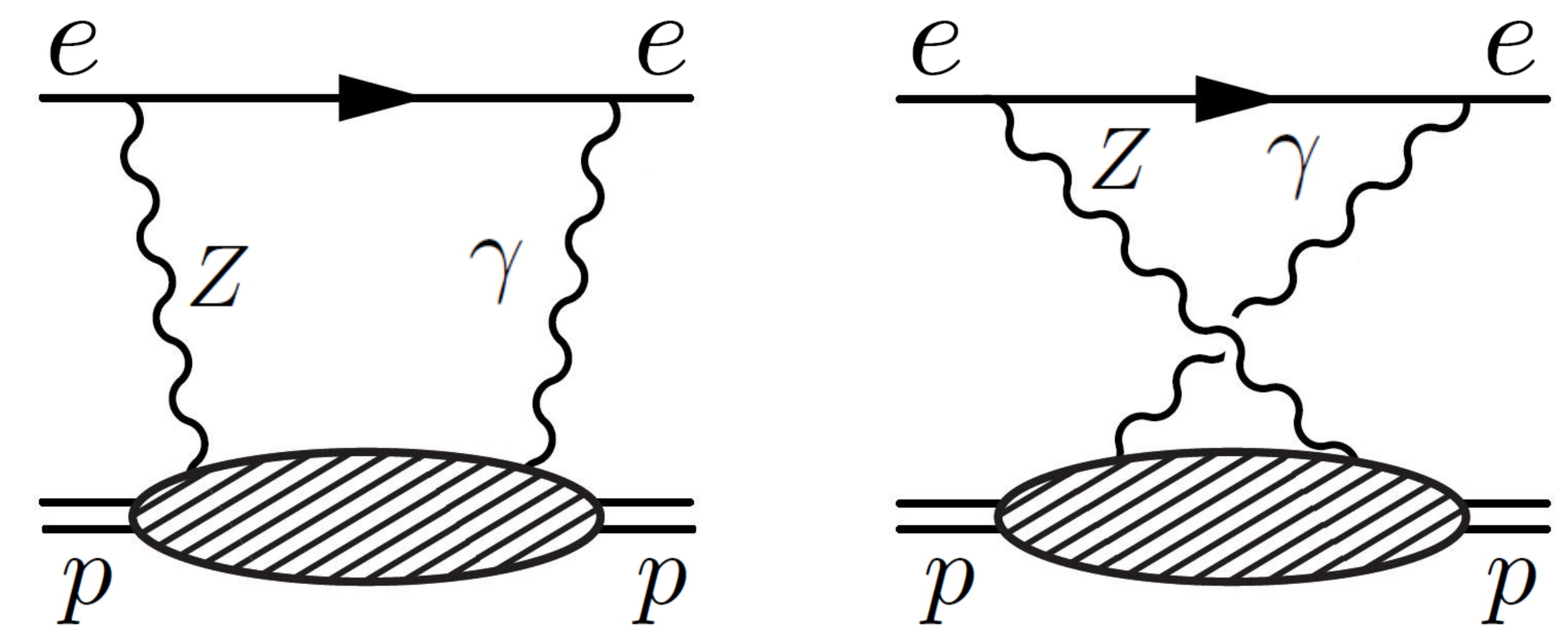}%\hfill
		\par\end{centering}
	\caption{
	The $\gamma Z$~box diagrams. Not shown are the two remaining
	diagrams with $\gamma$ and $Z$ interchanged.
	\label{fig:box}
	}
\end{figure}

Due to its accidental suppression at tree level, one needs to
carefully account for all the SM higher-order effects to $Q_W^p$
in order to properly translate the high-precision experimental
measurements into a determination of the weak mixing angle and
constraints on BSM parameters. Including
one-loop electroweak radiative corrections, the proton
weak charge reads~\cite{Erler:2003yk},
\begin{eqnarray}
Q_W^p
&=&
(1+\Delta\rho+\Delta_e)(1-4s_W^2(0)+\Delta_e')
\nonumber\\
&&
+\Box_{WW}+\Box_{ZZ}+\Box_{\gamma Z}(0).
\end{eqnarray}
Among them, the quantity $\Box_{\gamma Z}$ which denotes the
contribution from the $\gamma Z$~box diagram (see Fig.~\ref{fig:box})
contains sensitivity to physics at the hadronic scale where
perturbative calculations are unreliable, inducing a
large theoretical uncertainty. It is in general a function of $E$,
and can be split into two terms,
\begin{equation}
\Box_{\gamma Z}(E)
= \Box_{\gamma Z}^V(E)+\Box_{\gamma Z}^A(E),
\end{equation}
where the superscript $V(A)$ denotes the contribution from the
vector (axial-vector) weak neutral current on the hadron side.
The axial box is non-zero at $E=0$, but suppressed by the small
electron weak vector coupling $v_e = -(1 - 4s_W^2)$. On the other
hand, the vector box is not suppressed by any small coefficient,
but is exactly zero at $E=0$.

Earlier studies of $\Box_{\gamma Z}$~\cite{Marciano:1982mm,
Marciano:1983ss,Bardin:2001ii} assume small $E$ and hence are
sensitive only to the axial box. However, a re-analysis in 2009
based on a dispersion relation~\cite{Gorchtein:2008px} revealed a
steep energy-dependence in $\Box_{\gamma Z}^V(E)$ which was not
previously accounted for. This finding stimulated a large number
of follow-up studies on the vector $\gamma Z$ 
box~\cite{Sibirtsev:2010zg,Rislow:2010vi,
Gorchtein:2011mz,Hall:2013hta,Gorchtein:2015qha,Hall:2015loa,
Gorchtein:2015naa} as the latter was found to play a dominant role
in the extraction of the proton weak charge in the Qweak experiment
which took data at $E=1.165$ GeV~\cite{Androic:2018kni}. 
These studies consistently concluded that the theoretical
uncertainty in $\Box_{\gamma Z}^V(E)$ increases with $E$, which
also motivated future measurements of $Q_W^p$ (specifically, the P2
experiment) to be performed with lower beam energy.

%%%%%%%%%%%%%%%%%%%%%%%%%%%%%%%%%%%
%%%%%%%%%%%%%%%%%%%%%%%%%%%%%%%%%%%

Unlike its vector counterpart, there are much fewer follow-up
theoretical studies of $\Box_{\gamma Z}^A(E)$
\cite{Blunden:2011rd,Blunden:2012ty,Rislow:2013vta,Gorchtein:2015qha},
despite its becoming increasingly important at small $E$.
In particular, no attempt was made so far to relate the contributions
from multi-hadron intermediate states at small momentum transfers
below about 1~GeV$^2$
to experimental data, and all existing studies of these contributions
are based on {\it ad hoc} models where a systematic uncertainty
analysis is difficult or impossible. 

In this paper we present an updated study
of the axial $\gamma Z$~box by adopting a Regge parameterization
of the multi-hadron contributions with parameters fitted to inclusive
$\nu p/\bar{\nu}p$ scattering data. This technique was recently
introduced in the treatment of the axial $\gamma W$~box diagram in
neutron and nuclear $\beta$-decay~\cite{Seng:2018yzq,Seng:2018qru}.
We will show that by employing the neutrino data one is able to
further reduce the uncertainty in $\Box_{\gamma Z}^A$ quoted
in Ref.~\cite{Blunden:2011rd} by a factor of~2. Within the same
theoretical framework, we further discuss a contribution which
arises from the parity-odd component in the $\gamma\gamma$~box.
It possesses the same theoretical structure as $\Box_{\gamma Z}^A$
and is therefore inseparable from the latter. Using the current
constraints on the nucleon anapole moment, we provide an estimate
on the additional theory uncertainty induced by this term.

%\bigskip
%%%%%%%%%%%%%%%%%%%%%%%%%%%%%%%%%%%%%%%%%%%%%%%%%%%%%%%%%%%%%%%%%%%%%
\section{Dispersive representation of $\Box_{\gamma Z}^A$}
\label{sec:input}

We start from the dispersive representation of 
$\Box_{\gamma Z}^A$ derived in Ref.~\cite{Blunden:2011rd} 
for the case of electron scattering in the forward direction,
\begin{eqnarray}
\Box_{\gamma Z}^A(E)
&=&
\frac{2}{\pi} \int_0^\infty dQ^2
\frac{v_e(Q^2)\alpha(Q^2)}{Q^2(1+Q^2/M_Z^2)} \times
\nonumber\\
&& \int_0^1dxF_3^{\gamma Z}(x,Q^2)f(r, t^\prime),
\label{eq:integral}
\end{eqnarray}
where
\begin{equation}
f(r, t^\prime) =
\frac{1}{t^{\prime 2}} \mathrm{Re}\left[
\ln\left(1-\frac{t^{\prime 2}}{r^2}\right)
+ 2 t^\prime \tanh^{-1}\left(\frac{t^\prime}{r}\right)\label{eq:frt}
\right]
\end{equation}
with $x = Q^2/(W^2-M^2+Q^2)$, $r = 1 + \sqrt{1+4M^2x^2/Q^2}$,
$t^\prime= 4MEx/Q^2$, and $M$ the nucleon mass. We will take into 
account that in the $\overline{\mathrm{MS}}$ scheme both $s_W^2$ 
and $\alpha$ are running, {\em i.e.},\ scale
dependent parameters. We calculate them as described in 
Refs.~\cite{Erler:2017knj,Erler:1998sy}, respectively. 
The spin-independent, parity-odd structure function $F_3^{\gamma Z}$
is defined through the hadronic tensor,
\begin{eqnarray}
&&
\frac{1}{4\pi}\int d^4x e^{iq\cdot x}
\left\langle p(p)\right|[J_{em}^\mu(x),(J_Z^\nu(0))_A]\left|p(p)\right\rangle
\nonumber\\
&=&
-\frac{i\varepsilon^{\mu\nu\alpha\beta}q_\alpha p_\beta}{2p\cdot q}
F_3^{\gamma Z}(x,Q^2),
\end{eqnarray}
where $p$ is the 4-momentum of the incoming proton,
$q$ the 4-momentum transfer with $q^2 = - Q^2$, and $\varepsilon^{0123}=-1$.
The electromagnetic and weak neutral currents read,
\begin{eqnarray}
J_{em}^\mu =
\sum_{q}e_q\bar{q}\gamma^\mu q,\;\;\;\;
&&
J_Z^\mu =
(J_Z^\mu)_V + (J_Z^\mu)_A,
\nonumber \\
(J_Z^\mu)_V =
\sum_q g_V^q \bar{q}\gamma^\mu q,\;\;\;\;
&&
(J_Z^\mu)_A =
\sum_q g_A^q \bar{q}\gamma^\mu\gamma_5 q,
\end{eqnarray}
with $e_u=2/3$, $e_d=e_s=-1/3$, $g_V^q=2(I_{3,L}^q-2e_qs_W^2)$, and
$g_A^q=-2I_{3,L}^q$. In particular,
when $E=0$ one finds the simplification~\cite{Seng:2018qru},
\begin{equation}
\Box_{\gamma Z}^A(0)
=
\frac{3}{2\pi}\int_0^\infty \frac{dQ^2}{Q^2}
\frac{v_e(Q^2)\alpha(Q^2)}{1+Q^2/M_Z^2}M_3^{\gamma Z}(1,Q^2),
\end{equation}
where
\begin{equation}
M_3^{\gamma Z}(1,Q^2) =
\frac{4}{3}\int_0^1dx\frac{2r-1}{r^2}F_3^{\gamma Z}(x,Q^2)
\end{equation}
is the first Nachtmann moment of $F_3^{\gamma Z}$
\cite{Nachtmann:1973mr,Nachtmann:1974aj}.

%%%%%%%%%%%%%%%%%%%%%%%%%%%%%%%%%%%
\begin{figure}[t]
	\begin{centering}
		\includegraphics[width=0.9\linewidth]{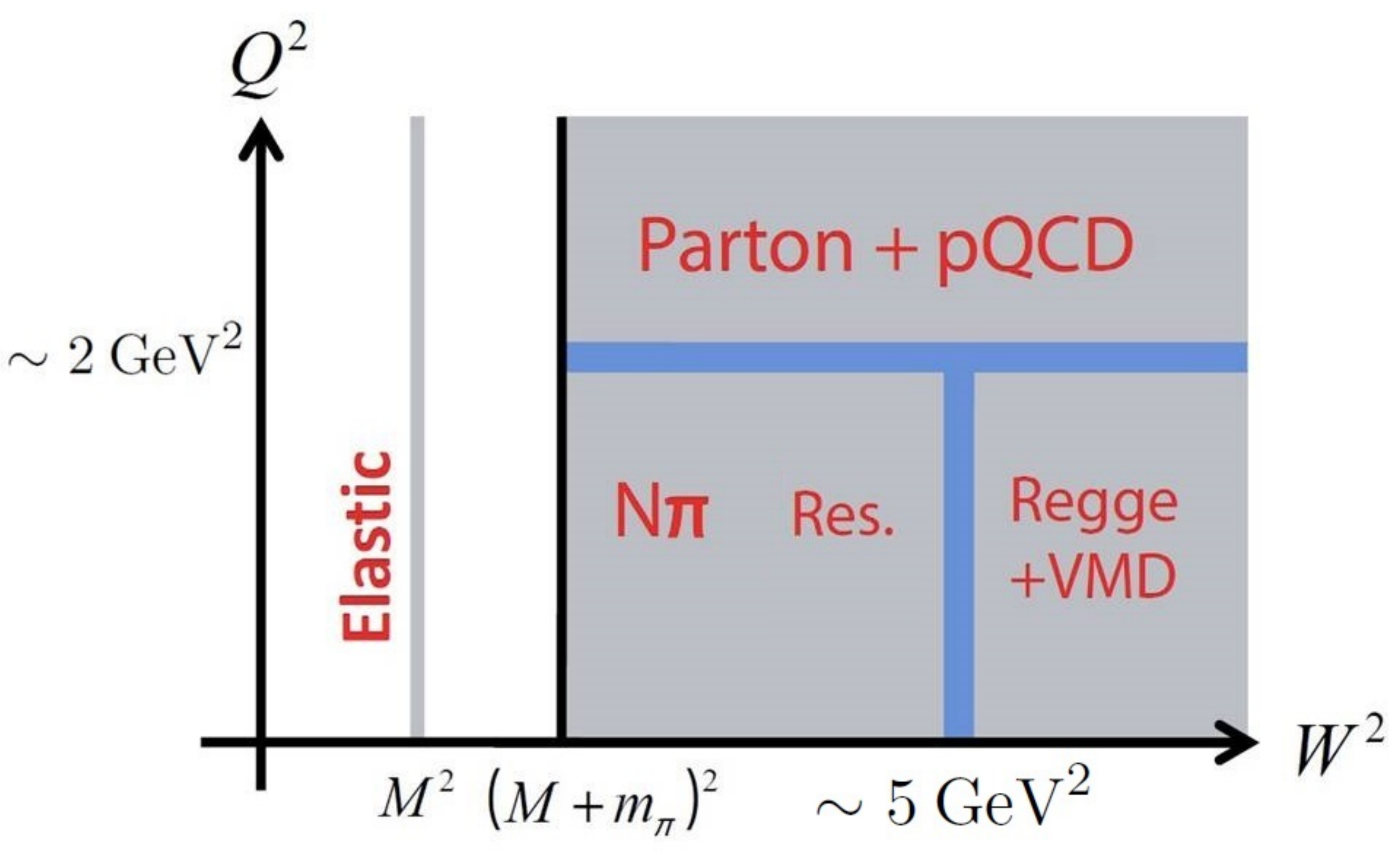}
		\par\end{centering}
	\caption{
	Dominant physics that enter $F_3^{\gamma Z}$ in different
	regions in the $W^2-Q^2$ plane.
	\label{fig:phase}
	}
\end{figure}
%%%%%%%%%%%%%%%%%%%%%%%%%%%%%%%%%%%

It is important to identify the dominant contributions to the
structure function $F_3^{\gamma Z}$ at different $Q^2$ and
$W^2=(p+q)^2$, as we summarize in Fig.~\ref{fig:phase}.
One can write,
\begin{equation}
F_3^{\gamma Z}
=
F_{3,\mathrm{el}}^{\gamma Z}
+ F_{3,\mathrm{inel}}^{\gamma Z},
\end{equation}
separating the elastic (sometimes called Born) contribution that
represents an isolated peak at $W^2=M^2$, and all the inelastic
contributions that start to emerge above the single-pion production
threshold $W^2 = (M+m_\pi)^2$. The latter can be further subdivided
in $Q^2$:
\begin{equation}
F_{3,\mathrm{inel}}^{\gamma Z}
=
\begin{cases}
F_{3,\pi N}^{\gamma Z}
+ F_{3,\text{res}}^{\gamma Z}
+ F_{3,\mathbb{R}}^{\gamma Z}
& \hspace{1pt} (Q^2 < 2\text{ GeV}^2)
\\[2mm]
F_{3,\mathrm{DIS}}^{\gamma Z} & \hspace{1pt} (Q^2 > 2\text{ GeV}^2)
\end{cases}
\end{equation}
First, above $Q^2 \simeq 2$ GeV$^2$, in the deep-inelastic
scattering (DIS) regime, the structure functions are well-described
by the parton model including corrections from perturbative quantum
chromodynamics (pQCD). This is corroborated by the observation
that the pQCD-corrected Gross-Llewellyn-Smith (GLS) sum rule
\cite{Gross:1969jf} is well satisfied at $Q^2>2$ GeV$^2$ in
$\nu p/\bar{\nu}p$ scattering experiments
\cite{Kataev:1994ty,Kim:1998kia}. 

In contrast, at lower $Q^2$
the effective degrees of freedom are hadrons. 
It was pointed out recently that nonperturbative 
contributions can in principle be obtained from lattice
QCD~\cite{Seng:2019plg}, but before such calculations are
carried out and the respective uncertainties are well understood, 
we have to rely on a model constrained by the experimental input. 

The lowest inelastic hadronic state, the $N\pi$ continuum, starts contributing above its production threshold. 
In the range $1.5~\mathrm{GeV}^2 < W^2 < 5~\mathrm{GeV}^2$ one observes
nucleon resonances.  
Above the two-pion threshold at $(M+2m_\pi)^2$ multi-hadron states emerge
which become dominant at large $W^2$. The Regge exchange picture provides an economical 
 description of these contributions which we smoothly continue down to the 
 two-pion threshold, as described in the following. We furthermore use the Vector Meson 
 Dominance (VMD) picture to extend the Regge description to moderate values of $Q^2$.
 We label this contribution by the subscript ``$\mathbb{R}$''. 
Correspondingly, we obtain the $\gamma Z$-box correction as 
a sum of five contributions,
\begin{equation}
\Box_{\gamma Z}^A(E)=\sum_{i}\Box_{\gamma Z}^{A,\,i}(E),
\end{equation}
with $i=$ el, $\pi N$, res, $\mathbb{R}$, and DIS. Each contribution 
is computed by inserting the corresponding 
$F_{3,i}^{\gamma Z}$ into Eq.~(\ref{eq:integral}) and integrating 
over its support in $W$ and $Q^2$. 

The elastic contribution is given in terms of  the proton magnetic Sachs form factor $G_M^p$ 
and the axial form factor $G_A$,
\begin{equation}
F_{3,\mathrm{\rm el}}^{\gamma Z}(x,Q^2)
=
- G_M^p(Q^2) G_A(Q^2) \delta(1-x).
\end{equation}
In the numerical evaluation of this part
we adopt recent data on these form factors as given in
Refs.~\cite{Ye:2017gyb,Bhattacharya:2011ah}.

The DIS region contribution to the box diagram can be represented as  
an infinite sum of odd Mellin moments of $F_3^{\gamma Z}$ upon expanding the function $f(r,t')$ in Eqs.~(\ref{eq:integral},\ref{eq:frt}) 
in powers of $x^2M^2/Q^2$~\cite{Blunden:2011rd}. None of the even Mellin moments of $F_3^{\gamma Z}$ appear due to the symmetry of the integrands in $x$.
The result is almost $E$-independent.
Furthermore, the size of contributions from the third and higher moments to $\Box_{\gamma Z}^A$ is about $10^{-5}$. 
It is two orders of magnitude below that from the first Mellin moment and has little impact on the final result. 
The first Mellin moment is independent of the details of the parton distribution functions
and its pQCD correction has previously been considered up to $\mathcal{O}(\alpha_s)$.
For completeness, we include here the full $\mathcal{O}(\alpha_s^4)$ expression~\cite{Baikov:2010je,Baikov:2010iw},
\begin{eqnarray}
\int_0^1dxF_{3,\mathrm{DIS}}^{\gamma Z}(x,Q^2) =&& \\  \nonumber
 \frac{5}{3}\left[1 - \frac{\alpha_s}{\pi} - c_2 \frac{\alpha_s^2}{\pi^2} -
c_3 \frac{\alpha_s^3}{\pi^3} - c_4\frac{\alpha_s^4}{\pi^4} \right] - && 
\frac{3}{2} \frac{\alpha_s^3}{\pi^3} \left[s_3 + s_4\frac{\alpha_s}{\pi}\right]. 
\end{eqnarray}
The first term on the right hand side~\cite{Baikov:2010je} represents the iso-singlet piece satisfying the polarized
Bjorken sum rule~\cite{Bjorken:1966jh}.
It receives only contributions from non-singlet (connected) diagrams.
The second term~\cite{Baikov:2010iw} contributes only to the iso-triplet piece by singlet (disconnected) diagrams
and enters the GLS~\cite{Gross:1969jf} sum rule.
The coefficients in numerical form are given by~\cite{Baikov:2010je,Baikov:2010iw},
\begin{eqnarray}
c_2 &=& 4.583 - 0.333\, n_f,
\nonumber\\
c_3 &=& 41.44 - 7.607\, n_f + 0.177\, n_f ^2,
\nonumber\\
c_4 &=& 479.4 - 123.4\, n_f + 7.697\, n_f^2 - 0.1037\, n_f^3,
\nonumber\\
s_3 &=& \phantom{479.4} - 0.413\, n_f,
\nonumber\\
s_4 &=& \phantom{479.4} - 5.802\, n_f + 0.2332\, n_f^2.
\label{eq:f3nlo-coeffs}
\end{eqnarray}
In the numerical evaluation, we use the running strong coupling
constant $\alpha_s$ provided by the Mathematica code
RunDec~\cite{Chetyrkin:2000yt} with $\alpha_s(M_Z)=0.118$,
and values for the number of flavors, $n_f$, changing by
one unit at quark mass thresholds as implemented in RunDec.
Integrating this expression with $Q^2_{\text{min}} = 2$~GeV$^2$
in Eq.~(\ref{eq:integral}) results in a contribution of 
$30.4\times 10^{-4}$ to
$\Box_{\gamma Z}^{A}$. We note that our $\mathcal{O}(\alpha_s^4)$
result differs from the $\mathcal{O}(\alpha_s)$ result by only about~2\%.
%so we expect the $\mathcal{O}(\alpha_s^4)$ corrections to be negligible. 
We have also checked that using $Q^2_{\text{min}} = $1~GeV$^2$ and 
the $\mathcal{O}(\alpha_s)$ expression we obtain
$32.9\times 10^{-4}$ in good agreement with Ref.~\cite{Blunden:2011rd}. 

Higher twist (HT) effects represent a source of the possible uncertainty in the lower 
$Q^2$ part of the DIS region, 2~GeV$^2<Q^2<5$~GeV$^2$. Since we have experimental 
data to compare with the pQCD prediction, we assume that the uncertainty due to HT
contributions should not exceed that of the data, which is $\sim$5\%. This uncertainty would be 
a systematical one, and we will use as the estimate 5\% of the integral of the DIS contribution 
over this $Q^2$-range. Numerically, it amounts to 1.3$\times10^{-5}$ which we include in the total error.

The $N\pi$ contribution is calculated within the framework of
chiral perturbation theory, but with pointlike electroweak vertex couplings 
replaced by the Dirac and axial nucleon form factors, which
suppresses their high-$Q^2$ contribution. Details of this 
calculation are described in Ref.~\cite{Seng:2018qru}. We assign a generous 
30\% uncertainty due to higher chiral orders. This uncertainty, however, does not affect the total error budget.

As for the resonances, 
our treatment is the same as in Ref.~\cite{Blunden:2011rd}. The 
only numerically relevant contribution is due to the $\Delta$-resonance, which
we calculate using the parameterization in 
Ref.~\cite{Lalakulich:2005cs}. As there are no assigned 
uncertainties for the values of the parameters, it is not possible to provide 
an error estimate for the $\Delta$-contribution to 
$\Box_{\gamma Z}^A$. This contribution is small, at the level of the 
overall uncertainty, and even a conservative 30\% uncertainty hardly modifies the total 
error. Moreover, we note here that the resonance and multi-hadron contributions 
are anti-correlated, as it is their sum that is constrained by measured experimental cross sections. 
A significant increase  in one will have to be compensated for by a reduction in the other, so that 
the overall effect is smaller than if this anti-correlation was na\"ively neglected. We therefore 
opt to exclude the resonance uncertainty from the total error. 

%%%%%%%%%%%%%%%%%%%%%%%%%%%%%%%%%%%%%%%%%%%%%%%%%%%%%%%%%%%%%%%%%%%%%
%\subsection{Relating $F_{3,\mathbb{R}}^{\gamma Z}$ to 
%$F_{3,\mathbb{R}}^{WW}$ by isospin symmetry}
%\label{sec:isospin}

%%%%%%%%%%%%%%%%%%%%%%%%%%%%%%%%%%%
\begin{figure}[t]
	\begin{centering}
		\includegraphics[width=0.4\linewidth]{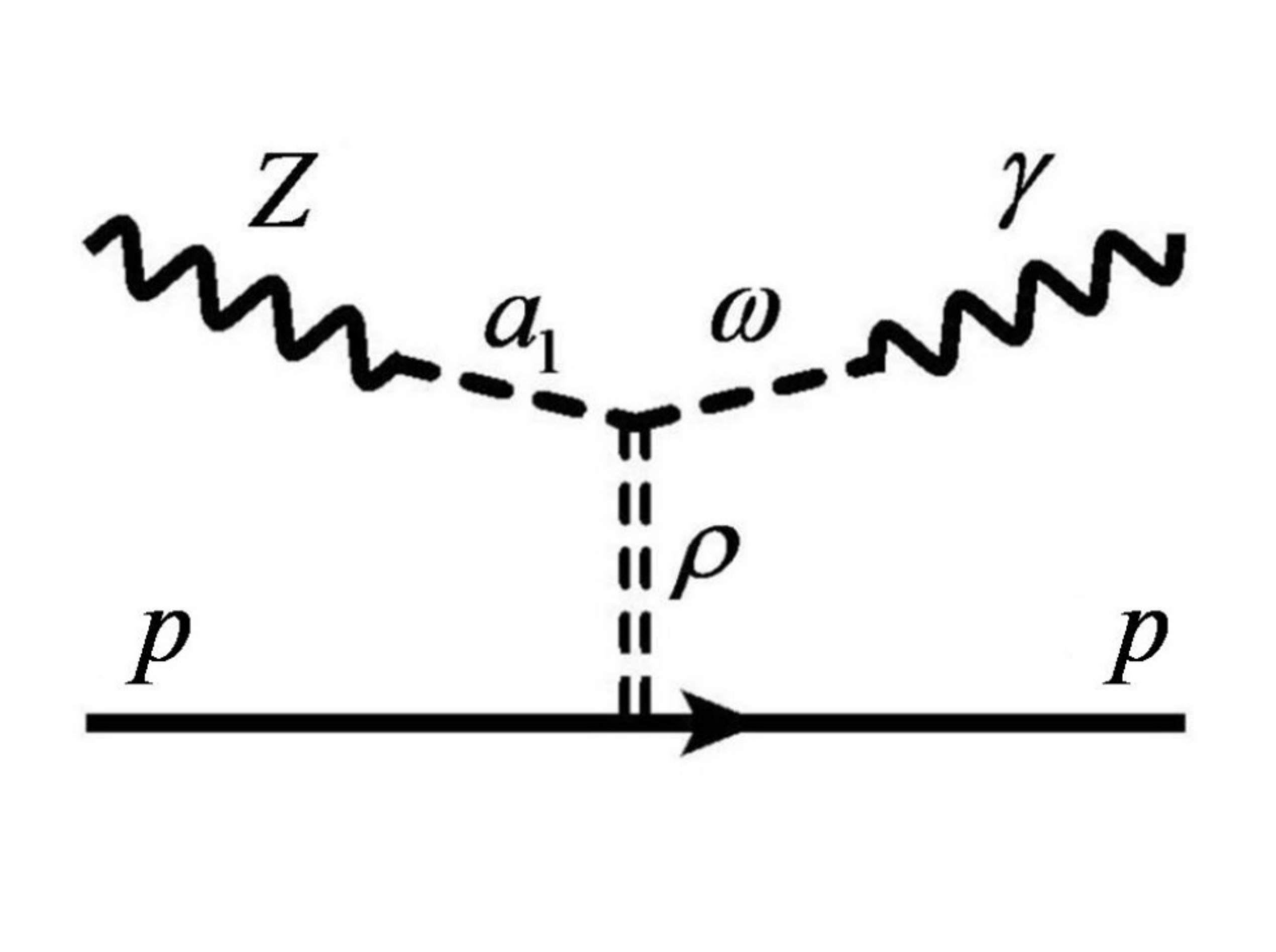}
		\includegraphics[width=0.4\linewidth]{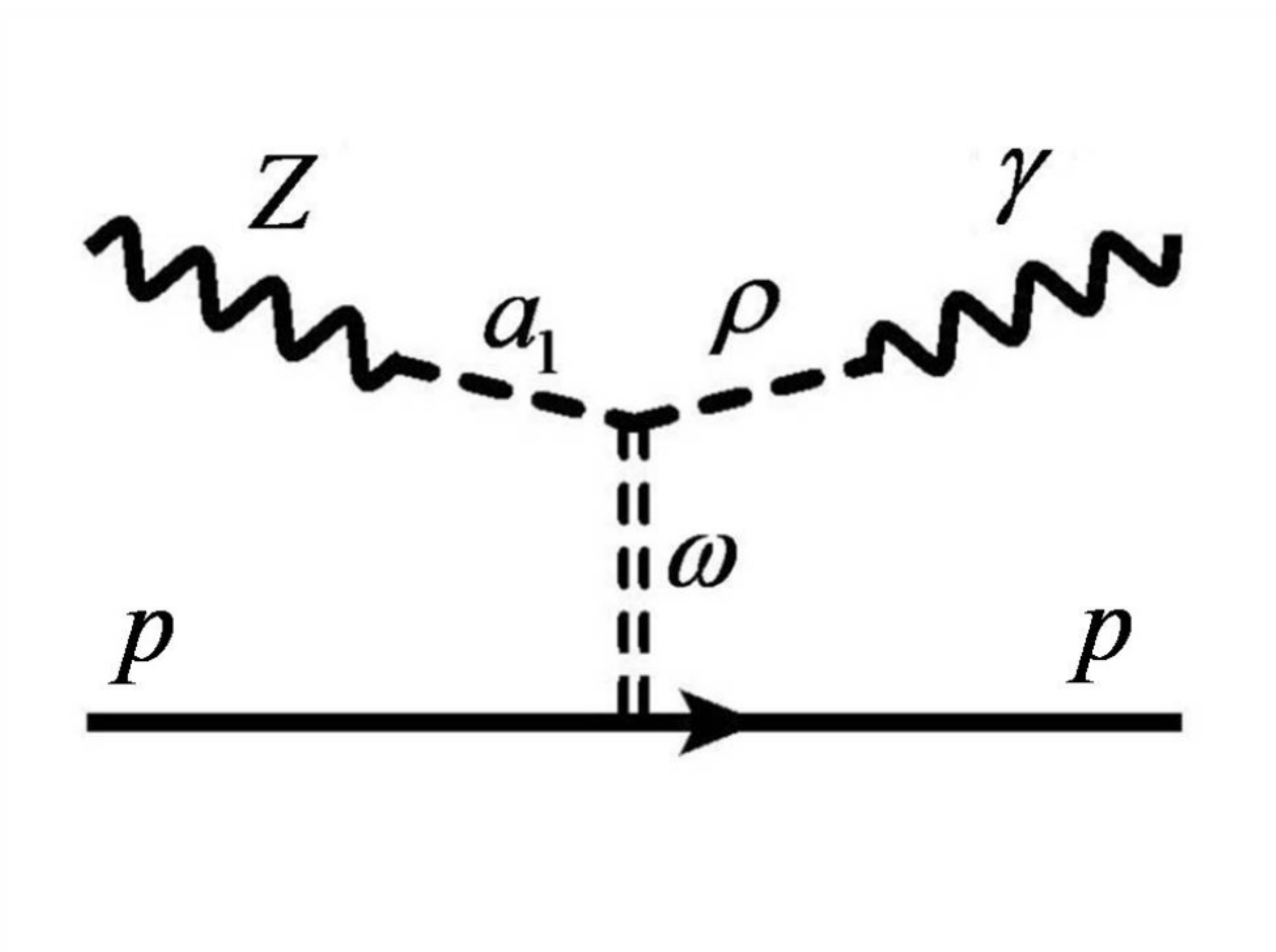}
		\par\end{centering}
	\caption{
	Regge exchange diagrams giving rise to the iso-singlet,
	$F_{3,\mathbb{R}(0)}^{\gamma Z}$, and iso-triplet,
	$F_{3,\mathbb{R}(1)}^{\gamma Z}$, parts of
	$F_{3,\mathbb{R}}^{\gamma Z}$, respectively.
	\label{fig:Regge}
	}
\end{figure}
%%%%%%%%%%%%%%%%%%%%%%%%%%%%%%%%%%%

The treatment of multi-hadron intermediate states requires more 
care. Following Refs.~\cite{Seng:2018yzq,Seng:2018qru},
we establish the correspondence between $F_{3}^{\gamma Z}$ and its
charged current counterpart $F_3^{\nu p+\bar{\nu}p}$ by isospin symmetry.
The electromagnetic current can be decomposed into iso-singlet
and iso-triplet components,
\begin{equation}
J_{\text{em}}^\mu
=
J_{\text{em}}^{\mu(0)} + J_{\text{em}}^{\mu(1)}.
\end{equation}
Likewise, we can also separate the structure function $F_3^{\gamma Z}$ into iso-singlet and
iso-triplet parts, $F_3^{\gamma Z} = F_{3(0)}^{\gamma Z} + F_{3(1)}^{\gamma Z}$,
where the axial part of the $Z$ current is purely iso-vector. 
Upon neglecting strange quarks, we
find through an isospin rotation that the $(I=1)\otimes(I=1)$ isospin channel,
$F_{3(1)}^{\gamma Z}$, equals half of the structure function
$F_3^{\nu p+\bar{\nu}p}$ that accounts for the difference of the
cross sections for inclusive $\nu p$ and $\bar{\nu}p$
scattering~\cite{Onengut:2005kv}.
We thus obtain the model-independent expression,
\begin{equation}
F_3^{\gamma Z}
=
\frac{1}{2}F_3^{\nu p+\bar{\nu}p}
+ F_{3(0)}^{\gamma Z}.
\end{equation}
There are two advantages in doing so, as (a) data exist for
$F_3^{\nu p+\bar{\nu}p}$ at moderate $Q^2$ where a
first-principle theory description is not available,
and (b) $J_{\text{em}}^\mu$ is predominantly iso-vector,
so that the contribution from $F_{3(0)}^{\gamma Z}$ is small.
Therefore, utilizing neutrino data, one obtains a
model-independent determination of the dominant iso-triplet
piece, limiting the model dependence to the small iso-singlet piece. 
Apart from the use of isospin symmetry, our treatment  
differs from that in Ref.~\cite{Blunden:2011rd} in that we 
account for continuous backgrounds already starting at the two-pion
threshold $W^2=(M+2m_\pi)^2$. 

%%%%%%%%%%%%%%%%%%%%%%%%%%%%%%%%%%%
\begin{figure}[t]
%	\begin{centering}
\vspace{-0.5cm}
		\includegraphics[width=1\linewidth,right]{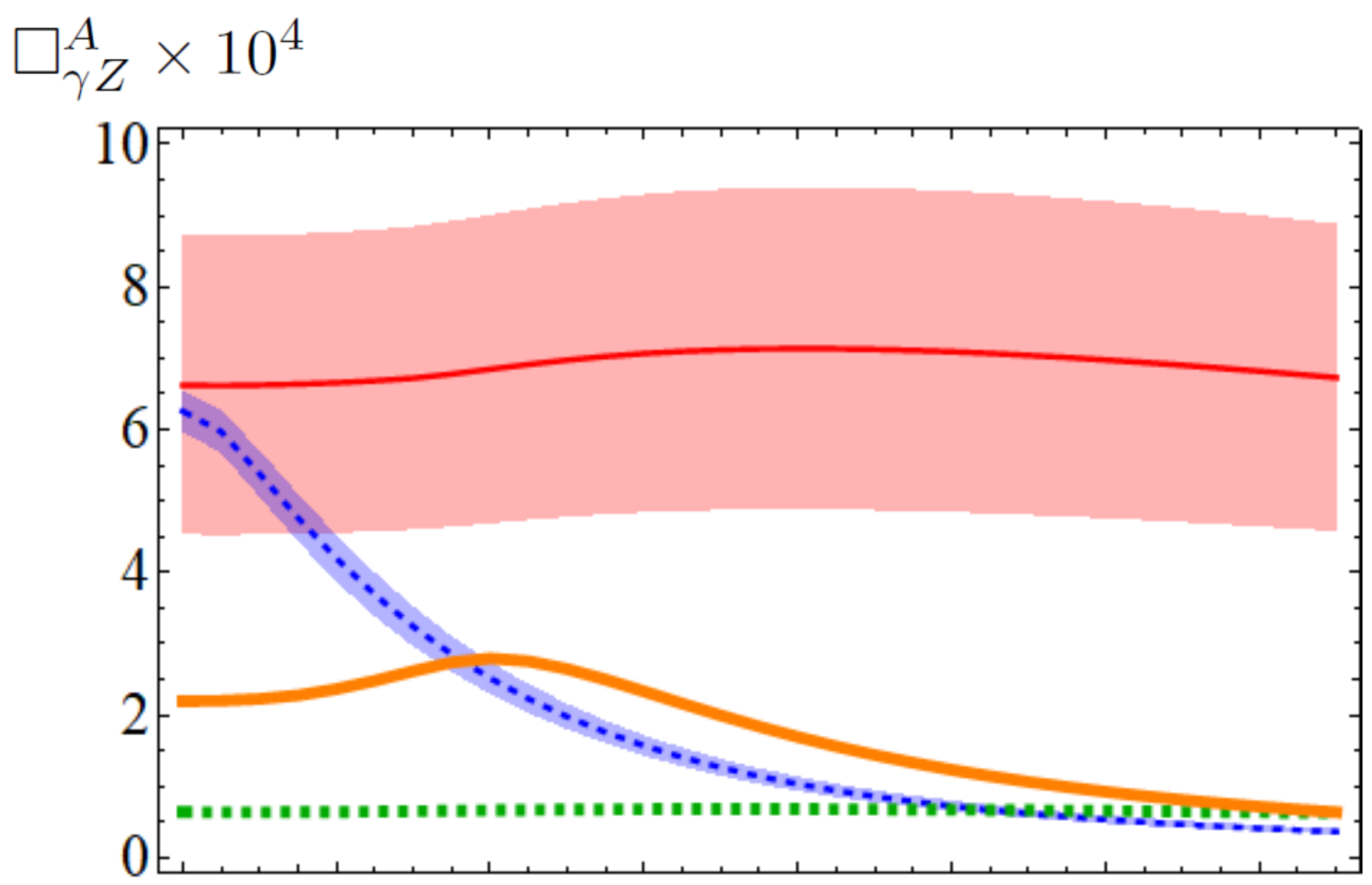}
		%\vspace{0.5cm}
		\includegraphics[width=0.965\linewidth,right]{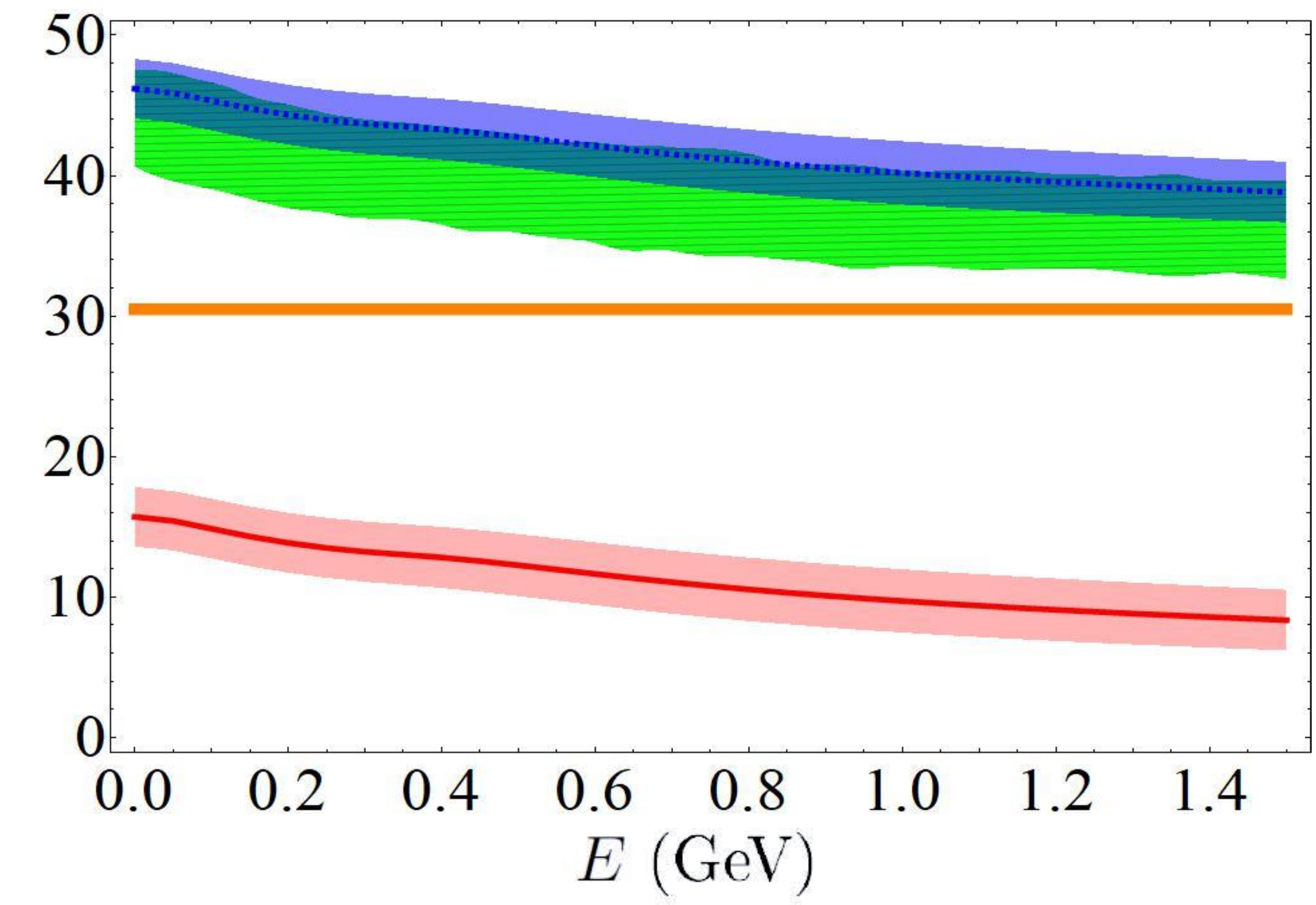}
%		\par
%		\end{centering}
\vspace{-0.7cm}
	\caption{
	Upper panel: summary of all non-DIS contributions to $\Box_{\gamma Z}^A$.
	Elastic (blue shaded region with dashed central line), resonance
	(orange solid line), $N\pi$ (green dashed line) and Regge (red
	shaded region with solid central line) contributions. 
	Lower panel: the DIS contribution (orange line), the sum of all other
	contributions (red shaded region with solid central line), and
	the total (blue shaded region with dashed central line).
	The result of Ref.~\cite{Blunden:2011rd} with its uncertainty band 
	is shown for comparison (green shaded area).
	\label{fig:summary}
	}
\end{figure}
%%%%%%%%%%%%%%%%%%%%%%%%%%%%%%%%%%%

Our parameterization is based on the leading Regge trajectory exchange picture
introduced in Refs.~\cite{Seng:2018yzq,Seng:2018qru} where the
reader can find more details. The respective diagrams are shown in Fig.~\ref{fig:Regge}.
In this picture and making use of the nearly exact degeneracy between the $\omega$ and $\rho$
mesons, the iso-singlet and iso-vector pieces are related to each other and to their charged current counterpart
in a simple way,
\begin{eqnarray}
F_{3,\mathbb{R}(0)}^{\gamma Z} =
\frac{f_{th}}{18}C_{WW}
\frac{m_\omega^2}{m_\omega^2+Q^2}
\frac{m_{a_1}^2}{m_{a_1}^2+Q^2}
\left(\frac{\nu}{\nu_0}\right)^{\alpha_0}
\label{eq:ReggeF3I0}\\
F_{3,\mathbb{R}(1)}^{\gamma Z} =
\frac{f_{th}}{2}C_{WW}\frac{m_\rho^2}{m_\rho^2+Q^2}
\frac{m_{a_1}^2}{m_{a_1}^2+Q^2}
\left(\frac{\nu}{\nu_0}\right)^{\alpha_0}
\nonumber\\
= \frac{1}{2}F_{3,\mathbb{R}}^{\nu p+\bar{\nu}p}, \hspace{123pt}
\label{eq:ReggeF3I1}
\end{eqnarray}
with $\alpha_0\approx 0.477$ the intercept of the $\omega/\rho$ trajectory,
$\nu=Q^2/2Mx$, and $\nu_0=1$~GeV a typical hadronic scale. 
The threshold function that suppresses the small-$W^2$ contribution was chosen in the form,
\begin{equation}
f_{th}=\Theta\left(W^2-W_{th}^2\right)\left(1-e^{(W_{th}^2-W^2)/\Lambda_{th}^2}\right),
\end{equation}
where $\Lambda_{th}=1$ GeV, and $W_{th}^2=(M+2m_\pi)^2$ is the two-pion threshold.
As in Ref.~\cite{Seng:2018yzq}, we assume that it is sufficient to take $C_{WW} = C_{WW}(Q^2)$ 
as linear, $C_{WW} = A_{WW}(1 + B_{WW}Q^2)$.
The parameter values $A_{WW} = 5.2 \pm 1.5$ and
$B_{WW} = 1.08^{+0.48}_{-0.28}$~GeV$^{-2}$~\cite{Seng:2018yzq}
have been obtained from a fit of the first Nachtmann moment
of $F_3^{\nu p+\bar{\nu}p}$ using data from CCFR~\cite{Kataev:1994ty,Kim:1998kia},
BEBC/Gargamelle~\cite{Bolognese:1982zd} and
WA25~\cite{Allasia:1985hw}, after subtracting the elastic,
$N\pi$, and resonance pieces. Using
Eqs.~\eqref{eq:ReggeF3I0} and \eqref{eq:ReggeF3I1}
in Eq.~\eqref{eq:integral} gives the Regge contribution
to $\Box_{\gamma Z}^A$.

Our results for the $\Box_{\gamma Z}^A$ are shown as function of energy in Fig.~\ref{fig:summary}. 
The upper panel displays all non-DIS contributions separately, while the lower panel shows their sum, 
the DIS contribution, the total, as well as a comparison with the previous evaluation of 
Ref.~\cite{Blunden:2011rd}. A significant overlap of the uncertainty bands indicates 
an excellent agreement between the two calculations. 
In our analysis the uncertainty is reduced by a factor of almost 2. 
%The total including the
%energy-independent DIS contribution is summarized in
%Fig.~\ref{fig:summary2}, where a comparison with the previous evaluation of 
%Ref.~\cite{Blunden:2011rd} is shown. We notice an excellent agreement of the two calculations,
%with the uncertainty significantly reduced in our analysis. 

%%%%%%%%%%%%%%%%%%%%%%%%%%%%%%%%%%%
%\begin{figure}
%	\begin{centering}
%		\includegraphics[width=1\linewidth]{summaryplot3}
%		\par\end{centering}
%	\caption{
%	Summary of the energy dependence of $\Box_{\gamma Z}^A$.
%	The DIS contribution (orange line), the sum of all other
%	contributions shown separately in Fig.~\ref{fig:summary}
%	(red shaded region with solid central line), and
%	the total (blue shaded region with dashed central line).
%	The result of Ref.~\cite{Blunden:2011rd} with uncertainty 
%	is shown for comparison (green shaded area).
%	\label{fig:summary2}
%	}
%\end{figure}
%%%%%%%%%%%%%%%%%%%%%%%%%%%%%%%%%%%

%%%%%%%%%%%%%%%%%%%%%%%%%%%%%%%%%%%
\begin{figure}[t]
\vspace{-0.5cm}
	\begin{centering}
		\includegraphics[width=1\linewidth]{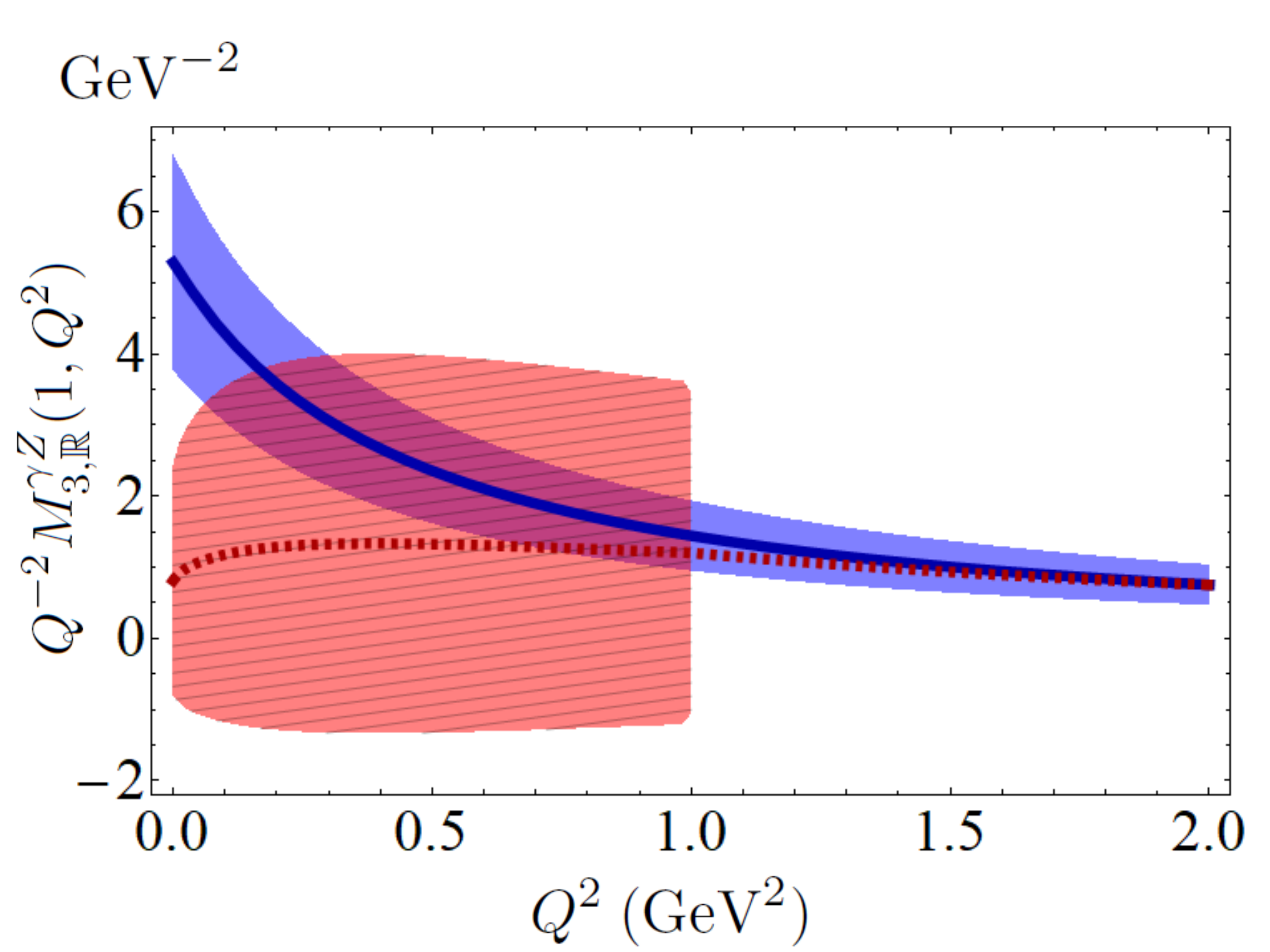}
		\par\end{centering}
\vspace{-0.2cm}
	\caption{
	Our Regge-VMD model result for the first Nachtmann moment of the structure function 
	$F_3^{\gamma Z}$ as a function of $Q^2$ (solid blue curve for the central value, and the shaded area around it for the uncertainty).
	For comparison, the low-$Q^2$ DIS contribution of Ref.~\cite{Blunden:2011rd} is shown for $Q^2<1$ GeV$^2$ (dashed red curve for the central value and red shaded area around for the uncertainty.)
	\label{fig:M1compare}
	}
\end{figure}
%%%%%%%%%%%%%%%%%%%%%%%%%%%%%%%%%%%

We visualize the source of this uncertainty reduction in Fig.~\ref{fig:M1compare} 
where we display our Regge contribution to the first Nachtmann moment $M_3^{(1)}(Q^2)/Q^2$ in comparison
with the result of the approach of Ref.~\cite{Blunden:2011rd}. That Ref. assumed no uncertainty beyond $Q^2=1$ GeV$^2$ 
where the dashed red curve is matched to DIS. Instead, we match the Regge parametrization to DIS at $Q^2=2$ GeV$^2$.
The total uncertainty of the $\Box_{\gamma Z}^A$ associated with each approach is obtained by integrating the respective 
uncertainty band in the displayed range. We see that  the use of data to determine the uncertainty of the 
dispersive calculation allows one to halve the associated theoretical error. The observed close agreement 
between the solid blue and dashed red curves in the range 1 GeV$^2<Q^2<2$ GeV$^2$ 
demonstrates that there is almost no sensitivity to the exact matching point in this range, and the respective error is contained 
in the shaded blue band. 

In Tab.~\ref{tab:table1} we list the individual contributions for the kinematics 
of the two relevant experiments, 
P2 at MESA with a beam energy of $E=155$~MeV, and QWeak at 
JLAB with $E=1.165$~GeV. Our updated analysis provides 
predictions for the axial $\gamma Z$~box with an uncertainty 
at the level of 5\%. For completeness, we quote here the previous evaluation by Ref.~\cite{Blunden:2011rd}:  
$\Box_{\gamma Z}^A(E=0)=44(4)\times 10^{-4}$ and 
$\Box_{\gamma Z}^A(E=1.165\,{\rm GeV})=37(4)\times 10^{-4}$. 

%%%%%%%%%%%%%%%%%%%%%%%%%%%%%%%%%%%
\begin{table}[t]
\begin{center}
\begin{tabular}{l|r r} 
  Contribution & \quad\quad $E=155$ MeV &  \quad\quad$E=1.165$ GeV \\ \hline
  elastic &4.7(3)\phantom{2.} &0.56(6)\phantom{.}   \\ 
  DIS &30.4(1)\phantom{2.} & 30.4(1)\phantom{2.}   \\
  $N \pi$ &0.6(2)\phantom{1.}  &0.7(2)\phantom{1.}   \\
  resonanes & 2.3(7)*\phantom{.} &\!1.0(3)*\phantom{.}   \\
  Regge &6.6(2.1) &7.0(2.2)   \\
  \hline
  \textbf{Total} & 44.6(2.1)&39.7(2.2) 
\end{tabular}
\end{center}
\caption{
Individual contributions to $\Box_{\gamma Z}^A(E)$ and associated uncertainties  for the 
P2 and QWeak experiments in units of $10^{-4}$. The star at the resonance 
contribution indicates that it is 100\% anticorrelated with the Regge contribution, see the discussion 
of the uncertainties in the text.
}
\label{tab:table1}
\end{table}
%%%%%%%%%%%%%%%%%%%%%%%%%%%%%%%%%%%

%%%%%%%%%%%%%%%%%%%%%%%%%%%%%%%%%%%%%%%%%%%%%%%%%%%%%%%%%%%%%%%%%%%%%
\section{Parity-violating photon-hadron interaction}
\label{sec:PVgamma}

In the previous sections we computed the hadronic
structure-dependent one-loop corrections due to the exchange of
a $\gamma$ and a $Z$ boson between the electron and the proton.
This calculation requires information on the PV structure
function $F_{3}^{\gamma Z}$ which we obtained by relating it 
to its charged current partner $F_{3}^{WW}$ by isospin symmetry.
Since isospin breaking effects can be expected to be small, the
uncertainty in this calculation is dominantly experimental. 
Together with $\Box_{\gamma Z}^V$
\cite{Gorchtein:2008px,Sibirtsev:2010zg,Rislow:2010vi,
Gorchtein:2011mz,Hall:2013hta,Gorchtein:2015qha,Hall:2015loa,
Gorchtein:2015naa} and other one-loop corrections
\cite{Erler:2017knj} this completes the one-loop analysis for
the parity-violating part of the cross section.

To go beyond this result, one will need to perform challenging calculations of
two-loop effects. The
situation is expected to be particularly complicated when
addressing non-perturbative contributions. It is not evident
how a complete two-loop calculation at the hadronic level will
be viable since it will involve the analysis of time-ordered
products of three currents. While these hadronic
effects are not enhanced by large electroweak logarithms,
they may still be of importance at the $10^{-3}-10^{-4}$ level
which is the goal of the present analysis. Since it may not be possible 
to directly calculate these corrections, we aim at estimating the 
uncertainty which they may induce,
together with the shift in the central value of $Q_W^p$.

Of particular interest is a subclass of two-loop hadronic
corrections associated with the exchange of two bosons between
the electron and proton, and another boson exchanged within
the hadronic state. Since the loop integration is dominated
by momenta of a typical hadronic scale, $\ell \lesssim \Lambda_h
\sim 1$~GeV, one can conclude that the exchange of two heavy
bosons, $W$ or $Z$, will lead to corrections of 
${\cal O}(\alpha G_F\Lambda_h^2)$ which are negligible.
Furthermore, purely electromagnetic effects, {\em i.e.,}\ diagrams
with the exchange of photons only, cannot lead to a PV signature.
Thus, only diagrams with at least one $Z$ boson need to be
considered. Among these, the only significantly new contribution
arises from the exchange of two photons between the electron and
the hadronic system, with parity violation within the latter. 
Such PV effects may arise due to
mixing of hadronic states of equal spin and opposite parity.
The parity-odd effect in $2\gamma$-exchange thus induces a
parity-odd structure function $F_3^{\gamma\gamma}$ in the
$\gamma\gamma$~box. Its effect should be added to
$\Box_{\gamma Z}^A$ as it is indistinguishable experimentally.

\begin{figure}
	\begin{centering}
		\includegraphics[width=0.36\linewidth]{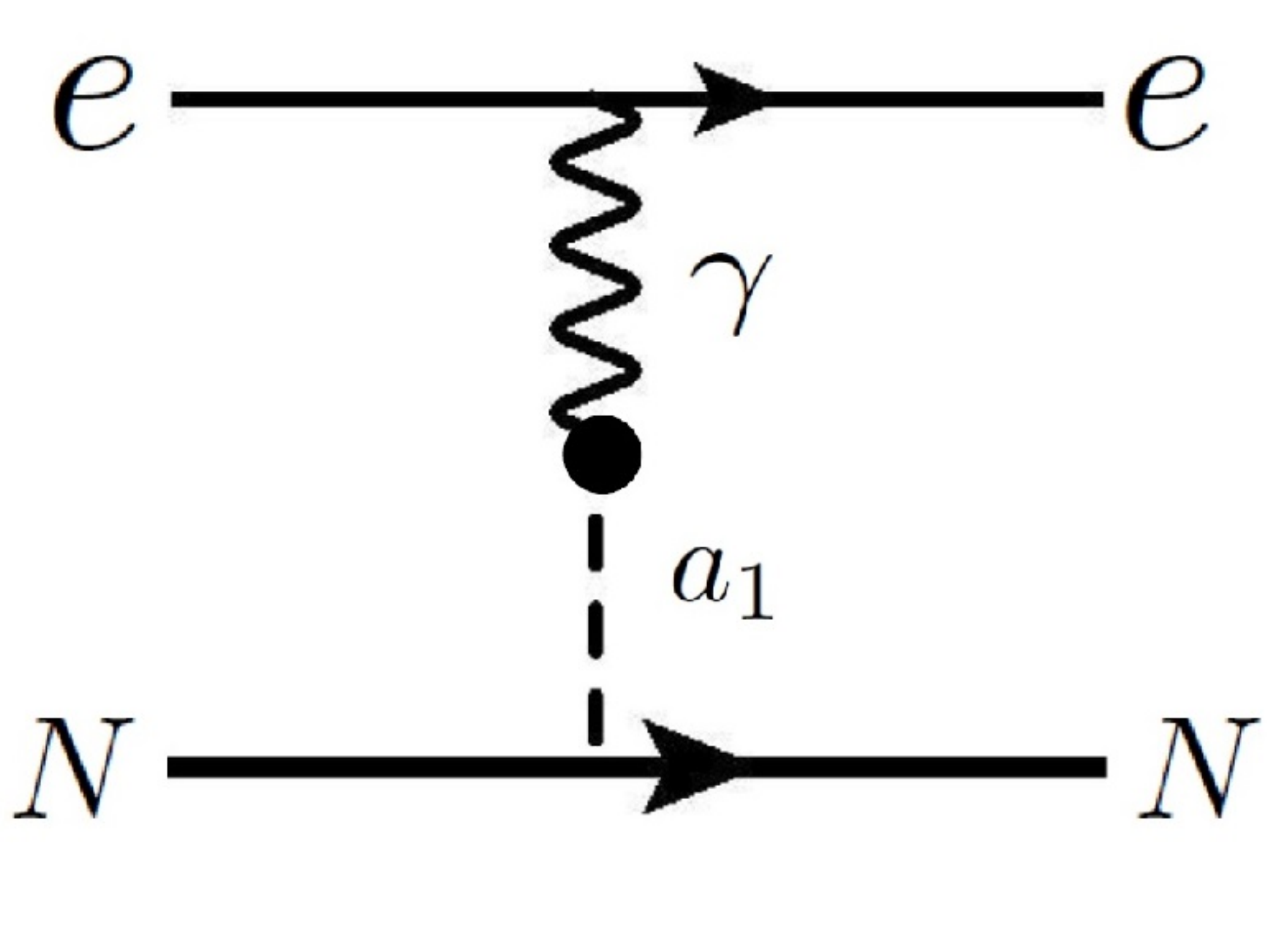}
		\qquad\quad
		\includegraphics[width=0.4\linewidth]{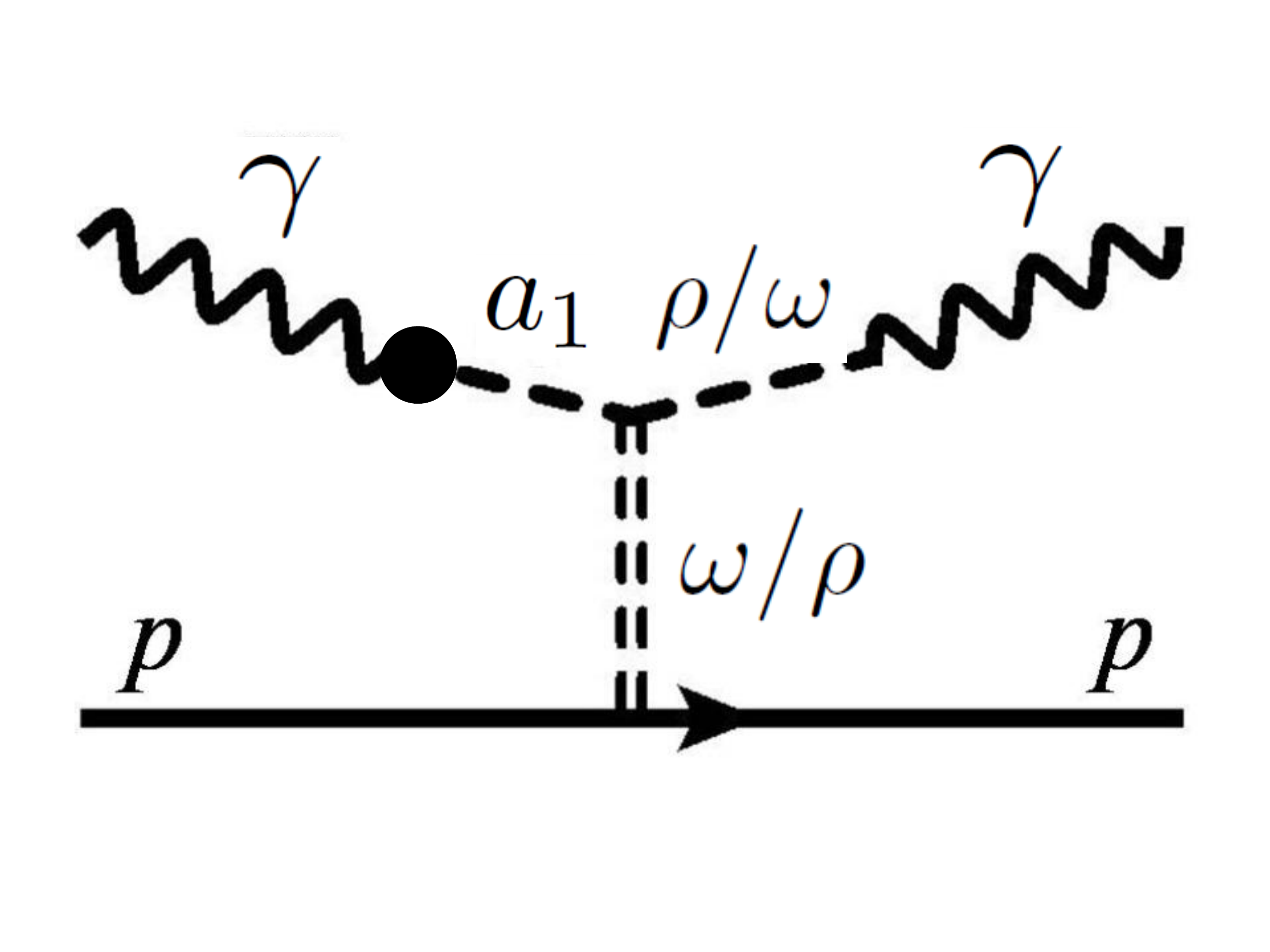}
		\end{centering}
	\caption{
	Left: effective description of the nucleon anapole moment.
	Right: Regge model description of PV in the forward
	$\gamma^*p$ Compton scattering amplitude.
	\label{fig:anapole}
	}
\end{figure}

In order to estimate this effect we use the picture in which
gauge bosons mix with vector and axial-vector mesons.
The photon mixes mostly with the vector mesons $\rho$,
$\omega$, and $\phi$. Hadronic PV interactions can induce
$a_1$--$\rho$ mixing, effectively leading to $a_1$--$\gamma$ mixing
(left panel of Fig.~\ref{fig:anapole}). Similarly, PV state
mixing in the iso-scalar channel, such as $h_1$--$\omega$,  gives
rise to $h_1$--$\gamma$ mixing. But due to
iso-vector dominance its numerical impact should be marginal
(at the 10\% level) and we disregard it. The Lagrangian 
describing the PV $a_1$--$\gamma$ interaction can be
written as
\begin{equation}
\mathcal{L}_{a_1\gamma}
=
\frac{e}{2}g_{a_1\gamma}F_{\mu\nu}a_1^{\mu\nu},
\label{eq:mixing}
\end{equation}
where $F^{\mu\nu}$ and $a_1^{\mu\nu}$ are the field strength
tensors of the photon and the $a_1$, respectively,
and $g_{a_1\gamma}$ is the PV coupling constant.

To study the elastic component of $F_3^{\gamma\gamma}$, 
we first recall that (neglecting strangeness and the iso-scalar axial coupling)
the nucleon matrix element of the axial weak neutral current takes the form,
\begin{equation}
\left\langle N\right|(J_Z^\mu)_A\left|N\right\rangle
=
g_A \, \bar{u}_N\gamma^\mu\gamma_5 \tau_3 u_N,
\end{equation}
where $g_A =  -1.27641(56)$ has been measured precisely in 
neutron beta decay~\cite{Markisch:2018ndu,Brown:2017mhw}. 
The inclusion of the so-called nucleon anapole moment
effectively shifts the apparent value of the nucleon axial charge seen in PV
electron scattering with respect to that in reactions involving
charged current interactions,
$g_A \rightarrow g_A+ \delta g_A^{ep}$.
Calculations based on hadronic parity violation  
in the framework of $SU(3)$ chiral perturbation theory~\cite{Kaplan:1992vj},
yield the result $\delta g_A^{ep} = 0.26(43)$~\cite{Zhu:2000gn}.
On the other hand, a global fit to PV electron scattering data not using
theory constraints returns an even higher value, 
%$g_A^{ep} = -0.62(63)$~\cite{Liu:2007yi,Gonzalez-Jimenez:2014bia}. 
$\delta g_A^{ep} = 0.66(63)$~\cite{Liu:2007yi,Gonzalez-Jimenez:2014bia}. 
These results are consistent with each other, as well as with zero,
but have large uncertainties impacting the error of the parity-violating asymmetry.
In the picture where the anapole moment arises due to the effective $a_1$--$\gamma$ mixing, 
as depicted in the left diagram of Fig.~\ref{fig:anapole}, the mixing strength $g_{a_1\gamma}$
is directly related to the value of $\delta g_A^{ep}$.
The elastic contribution to $F_3^{\gamma\gamma}$ is then obtained as~\cite{Gorchtein:2016qtl},
\begin{equation}
F_{3,\mathrm{el}}^{\gamma\gamma}
=
\frac{\delta g_A^{ep}}{g_A}F_{3,\mathrm{el}}^{\gamma Z}
= -(0.20\pm0.34) F_{3,\mathrm{el}}^{\gamma Z},
\label{eq:Bornshift}
\end{equation}
where the numerical estimate is based on the theory result~\cite{Zhu:2000gn}.

%%%%%%%%%%%%%%%%%%%%%%%%%%%%%%%%%%%
%%%%%%%%%%%%%%%%%%%%%%%%%%%%%%%%%%%

Here we note an analogous relation for the correction 
to the Regge contribution discussed in the previous section,
\begin{equation}
F_{3,\mathbb{R}}^{\gamma\gamma}
=
\frac{\delta g_A^{ep}}{g_A}F_{3,\mathbb{R}}^{\gamma Z},
\end{equation}
which can be shown straightforwardly.
This corresponds to a parity-odd effect which can be visualized
by the right diagram of Fig.~\ref{fig:anapole},
where a photon mixes with an $a_1$ through
Eq.~\eqref{eq:mixing}, and then interacts with the nucleon
through the exchange of a $\rho/\omega$ trajectory. 
Combining these two effects, we find
an effective shift of $\Box_{\gamma Z}^A$ at P2 energies given by,
\begin{eqnarray}
\Box_{\gamma\gamma}^{PV}&=&-(0.20\pm0.34)\Big(\Box_{\gamma Z,\rm el}^A+\Box_{\gamma Z,\mathbb{R}}^A\Big)\nonumber\\
&=&-2.3(3.8) \times 10^{-4}.
\end{eqnarray}
One should keep in mind that further hadronic contributions not
related to the anapole moment are possible, but their impact on the central
value and the respective uncertainty is well below the target accuracy~\cite{Gorchtein:2016qtl}.

%%%%%%%%%%%%%%%%%%%%%%%%%%%%%%%%%%%%%%%%%%%%%%%%%%%%%%%%%%%%%%%%%%%%%
\section{Summary and conclusions}
\label{sec:results}

We provided a thorough update of the $\Box_{\gamma Z}^A$ correction to PV electron-proton 
scattering. This calculation entails gathering all available information on the interference PV 
structure function $F_3^{\gamma Z}$ and its first Nachtmann moment over the full range of $Q^2$. 
The limiting cases of low and high $Q^2$ are governed by the elastic and DIS contributions, respectively.
While these two contributions are known with good precision, the interpolation between them requires 
modelling inclusive hadronic contributions in the non-perturbative regime. This interpolation is the source 
of the uncertainty of the calculation. Luckily, the contributions from the intermediate $Q^2$ range are rather 
small, making the resulting model dependence not critical. In the past, the interpolation was performed by an 
essentially ad hoc procedure. In this work we invoke the isospin symmetry that is known to hold to a good extent, 
and relate the neutral current interference structure function $F_3^{\gamma Z}$ to its charged current counterpart 
$F_3^{WW}$ for which experimental data from inclusive neutrino and antineutrino 
scattering are available. Even though these data are not very precise, this procedure allowed us to better constrain the 
interpolation between the low and high $Q^2$ regimes, leading to a factor 2 reduction in the resulting uncertainty. 
In particular, for the P2
beam energy of $E = 155$ MeV we obtain from Table~\ref{tab:table1},
\begin{equation}
\Box_{\gamma Z}^A = (44.6 \pm 2.1) \times 10^{-4},
\label{totalerror}
\end{equation}
%$30.5\times 10^{-4}$ (DIS)
%+ $4.7(3)\times 10^{-4}$ (elastic)
%+ $0.6\times 10^{-4}$ ($N\pi$)
%+ $2.3\times 10^{-4}$ (resonance)
%+ $6.6(2.1)\times 10^{-4}$ (Regge)
%= $$,
which is in good agreement with Ref.~\cite{Blunden:2011rd} but with the
uncertainty reduced by a factor of about two.
Including our result for $\gamma\gamma$ exchange with
hadronic PV results in the total correction,
\begin{equation}
\Box_{\gamma Z}^A + \Box_{\gamma\gamma}^{PV} = (42.3 \pm 4.3) \times 10^{-4}.
\end{equation}
This uncertainty is significantly larger than the one in Eq.~(\ref{totalerror}).
We note, however, that the P2 experiment~\cite{Becker:2018ggl} will in any case aim at reducing the 
uncertainty of the proton's anapole moment $\delta g_A^{ep}$ by a factor of four via a dedicated backward angle measurement.
The relevant uncertainty will then be the one in Eq.~(\ref{totalerror}),
while the small remaining one due to the induced PV photon-hadron interaction 
is 100\% correlated with, and needs to be added linearly to, the tree-level $g_A$ effect in the PV asymmetry.

Our calculation of the hadronic structure-dependent one-loop
corrections due to the exchange of a $\gamma$ and a $Z$ boson
between the electron and the proton has to be combined with
other already known one-loop corrections~\cite{Erler:2017knj,
Gorchtein:2008px,Sibirtsev:2010zg,Rislow:2010vi,
Gorchtein:2011mz,Hall:2013hta,Gorchtein:2015qha,Hall:2015loa,
Gorchtein:2015naa}.
This completes the one-loop analysis for the parity-violating
part of the cross section. 

Until a complete two-loop calculation
is performed, one should try to identify the leading two-loop
effects. We have done this for hadronic PV effects
entering through $\gamma\gamma$ exchange. Our result shows
that two-loop contributions to the PV asymmetry in electron
proton scattering are numerically relevant. It seems
less important now to work on further improvements of the
one-loop results; instead theoretical efforts should shift
to the calculation of two-loop effects. 
%\bigskip

\newpage

%%%%%%%%%%%%%%%%%%%%%%%%%%%%%%%%%%%%%%%%%%%%%%%%%%%%%%%%%%%%%%%%%%%%%
\noindent {\bf Acknowledgements:} 
This work was supported by the German-Mexican research collaboration Grant No. 278017 (CONACyT) and No. SP 778/4-1 (DFG).
CYS is supported in part by the DFG (Grant No.~TRR110) and the NSFC (Grant No.~11621131001) 
through the funds provided to the Sino-German CRC 110 ``Symmetries and the Emergence of Structure in QCD", 
and also by the Alexander von Humboldt Foundation through a Humboldt Research Fellowship. 
MG acknowledges support by Helmholtz Institute Mainz.
JE acknowledges support by PASPA (DGAPA--UNAM) and CONACyT Project No. 252167--F,
and is grateful for hospitality and support by the excellence cluster PRISMA$^+$ at JGU Mainz, 
as well as the Helmholtz-Institute Mainz.

%%%%%%%%%%%%%%%%%%%%%%%%%%%%%%%%%%%%%%%%%%%%%%%%%%%%%%%%%%%%%%%%%%%%%

%\bibliographystyle{unsrtnotitle}
\bibliographystyle{apsrev4-1}
\bibliography{axialgammaZ_ref}

\end{document}